%% file: main.tex
\documentclass[final,3p,fleqn]{elsarticle}
\input{preamble}

\begin{document}

%%%%%%%%%%%%%%%%%%%%%%%%%%%%%%%%%%%%%%%%%%%%%%%%%%%%
\begin{frontmatter}
\title{
Financial Performance and Innovation: Evidence From USA, 1998 - 2023
}
\author{Panteleimon Kruglov}
\author{Charles Shaw}
\address{This version: 16 March 2024}
%%%%%%%%%%%%%%%%%%%%%%%%%%%%%%%%%%%%%%%%%%%%%%%%%%%%
\begin{abstract}

This study explores the relationship between R\&D intensity, as a measure of innovation, and financial performance among S\&P 500 companies over 100 quarters from 1998 to 2023, including multiple crisis periods. It challenges the conventional wisdom that larger companies are more prone to innovate, using a comprehensive dataset across various industries. The analysis reveals diverse associations between innovation and key financial indicators such as firm size, assets, EBITDA, and tangibility. Our findings underscore the importance of innovation in enhancing firm competitiveness and market positioning, highlighting the effectiveness of countercyclical innovation policies. This research contributes to the debate on the role of R\&D investments in driving firm value, offering new insights for both academic and policy discussions.

\end{abstract}
\end{frontmatter}
%%%%%%%%%%%%%%%%%%%%%%%%%%%%%%%%%%%%%%%%%%%%%%%%%%%%

\JEL{E44, O31
}

%%%%%%%%%%%%%%%%%%%%%%%%%%%%%%%%%%%%%%%%%%%%%%%%%%%%
\newpage

\listoffigures
\listoftables
\newpage
\tableofcontents

\newpage
%%%%%%%%%%%%%%%%%%%%%%%%%%%%%%%%%%%%%%%%%%%%%%%%%%%%
\section{Introduction}
\label{sec:intro}

Research and development (R\&D) is critical for competitive advantage, yet its effect on firm value and the underlying mechanisms remain debated. This study investigates R\&D intensity (i.e. R\&D expenditure divided by revenue) — a measure of a firm’s commitment to innovation — and its impact on financial performance. We examine how innovation influences a firm's capital structure, suggesting a complex relationship where reduced R\&D intensity may lead to negative investor valuation adjustments due to anticipated competitive disadvantages or growth limitations.

Bridging the theoretical gap, our analysis explores economic mechanisms by which innovation affects firm value, including potential changes in investor expectations and financial indicators like the cost of capital and stock valuations. The study delves into the nuanced effects of innovation on firm valuation within the technology sectors of the S\&P 500, where innovation's importance is pronounced yet challenged by measurement difficulties. R\&D intensity, while a useful proxy, may not fully capture innovation's breadth across industries due to data availability and transparency issues.

Given the mixed evidence on the innovation-financial performance nexus, this research aims to clarify R\&D intensity's role in firm valuation. We consider both short-term impacts, such as immediate competitive disadvantages, and long-term effects like market irrelevance or technological obsolescence. This study contributes to the discourse on innovation in corporate strategy and investment decision-making, providing insights into the economic mechanisms at play and enhancing understanding of the complex relationship between innovation intensity and firm value.

%%%%%%%%%%%%%%%%%%%%%%%%%%%%%%%%%%%%%%%%%

The remainder of this study is organised as follows: 
Section \ref{sec:litreview} offers an overview of the existing literature; 
Section \ref{sec:data} details the data description; 
Section \ref{sec:identification} outlines the econometric methodology; 
Section \ref{sec:hypotheses} formulates the hypotheses;
Section \ref{sec:results} presents the results; 
Section \ref{sec:discussion} engages in a discussion of the empirical findings; 
and finally, Section \ref{sec:conclusion} concludes.

%%%%%%%%%%%%%%%%%%%%%%%%%%%%%%%%%%%%%%%%%%%%%%%%%%%%
\section{Background and Related Literature}
\label{sec:litreview}
The scholarly literature pertaining to the relationship between innovation and firm performance may be traced back to Joseph Schumpeter's influential paradigm, which posits that firm dynamics are shaped by the introduction of new innovations into the market, resulting in an evolutionary process. According to Schumpeter's theory of creative destruction, firms have the potential to stimulate economic development and growth by introducing innovative products and processes. This theory challenges the static equilibrium perspective by presenting a dynamic perspective on market evolution, which is influenced by the competitive interactions among diverse firms.

According to Schumpeter \cite{S12, S34}, the central focus of economic development is in invention, with entrepreneurship playing a crucial role in facilitating the transformation of innovation into overall economic growth. This viewpoint emphasises the interaction between incentives for innovation at the firm level and the overall growth of the economy, emphasising the government's responsibility in closing the divide when companies are hesitant to innovate or recruit scientific professionals \cite{S98}.

The task of integrating Schumpeter's concepts into a coherent theory has proven to be difficult. However, a significant change occurred in the 1980s with the introduction of modern Schumpeterian growth theory and evolutionary models, as demonstrated by Nelson et al, \cite{N85, NW82} and Dosi \cite{D83}. These models highlight the importance of the innovation process in driving productivity growth and economic development. They examine the differences in innovation patterns among different industries and companies.

Romer \cite{R90} and Lucas \cite{L93} subsequently developed endogenous growth models that recognised knowledge as a significant driver of growth, with a particular emphasis on its spillover effects. 

Recent scholarly investigations have broadened their scope to encompass the multifaceted nature of innovation and the intricate dynamics of financial metrics. In the literature, patents are often employed as a measure of innovation, based on the premise that inventors typically seek patents for their innovations \cite{lanjouw2001}. This association is especially pronounced in certain sectors, like the pharmaceutical industry, where a significant portion of innovations, approximately 80\%, are patented \cite{arundel1998}. One important reason why patents continue to be widely recognized as indicators of innovation is attributed to their correlation with firm performance \cite{bloom2002} and the accessibility of extensive national patent databases \cite{hasan2010}.

However, the adequacy of patents as a universal metric for innovation is not immune from being challenged, notably:
\begin{itemize}
    \item Griliches \cite{griliches1979} and Pakes \& Griliches \cite{pakes1980} point out the diverse nature and industry-specific aspects of patents, arguing that they should not be the sole indicators, replacing microeconomic measures such as a firm's productivity.
    \item Miller \cite{miller2013} suggests that a substantial proportion of patents, up to 60\%, may not genuinely contribute to innovation or drive growth and development.
    \item Cohen et al. \cite{cohen2001} observe that small and medium-sized enterprises often forego patent registration due to the disproportionate transaction costs compared to the benefits derived from innovation.
\end{itemize}

Nevertheless, the existing body of empirical research on the factors that drive firm performance indicates a lack of agreement, which might be attributed to the inadequate ability of financial and industrial markets to adequately identify and incentivize enterprises that demonstrate innovation and efficiency. As argued by Demirel and Mazzucato \cite{DM09}, the efficiency of market selection processes in recognising and motivating creative efforts is called into question by contemporary markets, as they face challenges in distinguishing between enterprises that are innovative and efficient, and those that are less innovative.

%\subsection{Impact of Recessionary Periods on Innovation and the Efficacy of R\&D Intensity as a Measure}
The influence of economic downturns, particularly recessionary periods, on corporate innovation is a critical aspect of our study. We specifically measure innovation through R\&D intensity, defined as the ratio of R\&D spending to total revenue. This measure provides a direct and quantifiable metric to assess a firm's commitment to innovation, especially during economic recessions.

%\subsubsection{Recessions and Innovation}

%\subsubsection{R\&D Intensity vs. Patent Data}
When considering the measure of innovation, R\&D intensity offers several advantages over alternative metrics like patent data. First, R\&D intensity is a broader measure that captures not just the outcome (as patents do) but the input and ongoing commitment of a firm towards innovation. It encompasses all forms of R\&D expenditure, irrespective of whether they result in patentable inventions or not.

Moreover, patent data, while valuable, can sometimes be a limited indicator of innovation. Not all R\&D efforts result in patents, and not all patents are commercially viable or indicative of meaningful innovation. In industries where the patenting culture is less prevalent, or in forms of innovation that are not patentable, R\&D intensity provides a more inclusive and representative measure of a firm's innovative activities.

However, it is also important to acknowledge that R\&D intensity has its limitations. It does not account for the efficiency or effectiveness of R\&D spending and may not directly correlate with successful innovation outputs. Therefore, while R\&D intensity is a robust measure of innovation input, it should ideally be complemented with other metrics, such as patent data or innovation outcomes, for a comprehensive assessment of a firm's innovation performance.

The impact of recessionary periods on innovation, as measured by R\&D intensity, underscores the sensitivity of innovation investments to economic conditions. While R\&D intensity is an effective measure of innovation input, especially during economic downturns, its utility is maximized when used in conjunction with other innovation metrics.

%\subsection{Patents as an Indicator of Innovation}

%\subsection{Complexity in Measuring Innovation and Financial Indicators}

The task of quantifying innovation and correlating it with financial indicators presents significant challenges, primarily due to the multifaceted nature of innovation. Innovation, often conceptualised as patent counts in empirical studies, extends beyond mere expenditure. It encompasses a spectrum of activities including product development, process innovation, and the implementation of novel business practices. These dimensions of innovation are not always directly measurable or accurately captured by financial expenditure on R\&D alone.

Moreover, the financial indicators commonly employed in empirical analyses, such as total assets, EBITDA, and firm size, are themselves subject to varying interpretations and implications. These metrics, while providing valuable insights into a firm’s financial health and capacity, do not always linearly translate into innovation outcomes. For instance, a firm's size may offer resource advantages for innovation, but could also entail bureaucratic complexities that stifle creative processes.

The complexity is further amplified when considering the temporal aspect of innovation and financial performance. The benefits of R\&D investments may not be immediately apparent and often materialise over extended periods. This lag can complicate the task of establishing direct cause-and-effect relationships between financial indicators and innovation outcomes.

Additionally, external factors such as economic conditions, market competition, and regulatory environments play a crucial role in shaping a firm's innovation activities and financial performance. These factors can introduce noise and confounders into the analysis, making it challenging to isolate the effects of internal financial indicators on innovation.

In light of these complexities, it is essential for empirical researchers to adopt a nuanced approach in measuring innovation and interpreting its relationship with financial indicators. This involves acknowledging the limitations of traditional metrics, considering a broader range of innovation indicators, and accounting for external influences and lag effects. Such an approach will enable a more accurate and comprehensive understanding of the dynamics between financial performance and innovation.

\section{Data}
\label{sec:data}
\subsection{Data Descrption}
Data for this study were sourced from the Financial Modelling Prep (FMP) API\footnote{ \url{https://site.financialmodelingprep.com/developer/docs/}}, encompassing a comprehensive timeframe of 100 quarters, extending from the second quarter of 1998 to the second quarter of 2023. The dataset comprehensively includes all entities listed in the S\&P 500 index, cutting across diverse industry classifications. This research focuses specifically on firms domiciled in the United States, deliberately excluding entities from European exchanges to maintain uniformity in reporting standards and avoid potential discrepancies. Environmental, Social, and Governance (ESG) data were procured from the SEC's EDGAR database. The panel data set comprises 506 firms, observed over 100 periods, yielding a total of 45,386 observations.

\subsection{Variables and their Definitions, 1998q2 - 2023q2}
We select key variables over the period from the second quarter of 1998 to the second quarter of 2023. 

\begin{itemize}
\item \textit{R\&D intensity}  "Research and development expenses" / "Revenue". This is our key measure of a firm's commitment to research and development. R \& D intensity is "normed" i.e. rescaled to a [0, 1] interval, for analytical consistency. 

\item \textit{ESG Score} is an aggregate measure reflecting a firm's susceptibility and response to long-term governance, social, and environmental challenges. FMP's ESG Ratings are based on a variety of data sources, including corporate sustainability reports, ESG research firms, and government agencies.

\item \textit{Total Assets}  = "Net income" / "Return on assets"\footnote{ since Return on assets = net income  /  total assets}

\item \textit{EBITDA} represents earnings before the deduction of interest, taxes, depreciation, and amortization, serves as a proxy for operating profitability. 

\item \textit{Recession} indicator is based on the National Bureau of Economic Research's (NBER) Recession Indicator (USRECQ).

\item \textit{GFC} NBER's indicator for the Global Financial Crisis (2007–2008). 

\item \textit{Firm Size} the natural logarithm of total assets.

\item \textit{Tangibility}  = "Tangible asset value" / "Total assets"

\item \textit{Tax Rate} is derived as the effective tax rate, computed from the income tax provision relative to pre-tax income. 

\item \textit{Dividend Yield} is the dividend per share as a proportion of share price. These variables align with established metrics in the literature on firm valuation.  Dividend Yield  = ("dividend paid" / "share number")  /  price

\item \textit{Graham Number}, rooted in the valuation methodology proposed by Graham (2003), signifies a threshold for stock value assessment, combining earnings per share and book value per share. 
\end{itemize}

\subsection{Dynamics of  R\&D intensity}

\begin{figure}
    \centering
\includegraphics[width=0.5\textwidth]{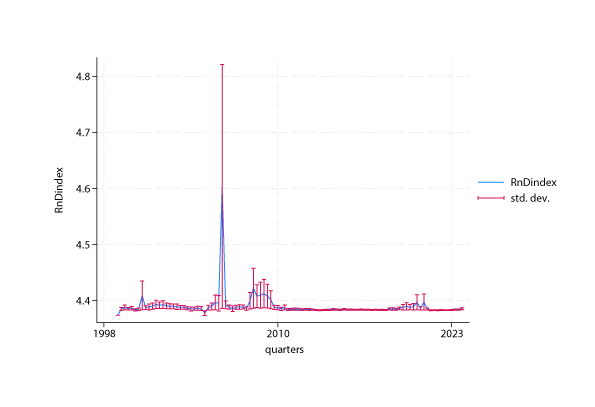}
    \caption{Evolution of R\&D index over time}
    \label{fig:rnd}
\end{figure}

Figure \ref{fig:rnd} illustrates the temporal evolution of the average R\&D intensity (R\&D Index) for firms within our sample. The graph reveals the fluctuating nature of R\&D investment over time, reflecting the dynamic prioritisation of research and development activities. The variations in R\&D intensity could be attributed to multiple converging factors:

\begin{itemize}
\item Fluctuating economic conditions which affect corporate profitability and discretionary spending on research and development.
\item The evolving technological landscape, which necessitates continual R\&D investment to maintain competitive advantage.
\item Periods of increased innovation, possibly aligned with technological breakthroughs or regulatory changes promoting research.
\end{itemize}

Moreover, the standard deviation of the R\&D Index appears to experience periods of volatility, suggesting changes in the distribution of R\&D investment across companies. This variability may be explained by:

\begin{itemize}
\item The diverse nature of R\&D strategies and capabilities across different industries within the S\&P 500.
\item The entry and exit of companies in the S\&P 500 index, which may have varying commitments to R\&D.
\item Shifts in industry focus, such as the transition towards digital technologies, which can lead to significant disparities in R\&D expenditure.
\end{itemize}

\subsection{Dynamics of Operational Profitability}

\begin{figure}
    \centering
\includegraphics[width=0.5\textwidth]{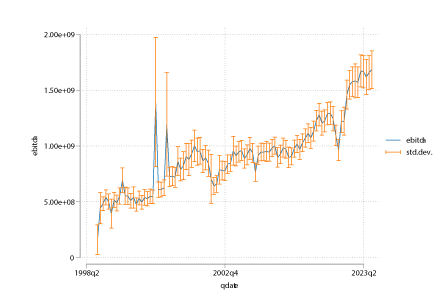}
    \caption{Evolution of EBITDA over time}
    \label{fig:EBITDA}
\end{figure}

Figure \ref{fig:EBITDA} elucidates the chronological progression of average EBITDA among S\&P 500 entities from 1998 through 2023. EBITDA—earnings before interest, taxes, depreciation, and amortisation—acts as a proxy for a firm’s underlying profitability absent financing, tax environments, and accounting depreciation policies. An analysis of the temporal data series reveals a discernible ascendant pattern in the mean EBITDA, indicative of an overall enhancement in operational earnings across the index constituents.

Several interwoven determinants may have catalysed this progressive increase in EBITDA:
\begin{itemize}
    \item An era of economic expansion characterising the terminal years of the twentieth century, potentially stimulating revenue augmentation and margin expansion.
    \item The advent and proliferation of digital commerce, which may have unveiled novel channels for revenue streams and operational efficiencies.
    \item A phase of deregulation in the financial domain, which conceivably streamlined the processes of capital procurement and thus invigorated corporate growth endeavours.
\end{itemize}

Concurrently, the graphical representation portrays a contraction in the standard deviation of EBITDA scores, inferring a decrease in disparity regarding operational profitability amongst the analysed corporations.

This proclivity towards uniformity might be attributable to several factors:
\begin{itemize}
    \item An ongoing consolidation trend within key sectors, culminating in a landscape populated by large-scale entities with analogous financial architectures.
    \item Incremental stringency in regulatory frameworks overseeing financial conduct, possibly curtailing the engagement in high-variance financial strategies.
    \item The intricate implications of global economic integration, necessitating a more sophisticated approach to cross-jurisdictional financial assessments.
\end{itemize}

While the ascent in average EBITDA and the contraction of its variability are observable phenomena, the underlying mechanisms warrant additional empirical scrutiny, which extends beyond the ambit of this discourse. The intricate interplay of these influences, particularly in relation to financial regulatory dynamics, underscores the complexity inherent in the corporate profitability narrative.

\subsection{Diagnostic Evaluations}
\subsubsection{Assessment of Multicollinearity}
In the preliminary stage of our regression diagnostics, we examined the Variance Inflation Factor (VIF) to discern the degree of multicollinearity among the independent variables. Multicollinearity complicates the isolation of individual variable effects and can engender estimates that lack precision and stability. A threshold VIF value above 10 typically signals substantial multicollinearity concerns. Our analysis yielded VIF values not exceeding 1.5 for any variable, denoting a satisfactory absence of multicollinearity within the model's architecture.

\subsubsection{Panel Data Stationarity: ADF Unit Root Test}
\label{sec:ADF}
The stationarity of panel data series was investigated utilising a Fisher-type unit-root test, predicated on the augmented Dickey–Fuller (ADF) framework. This test operates under the null hypothesis (H0) of pervasive non-stationarity across the panels. The alternative hypothesis posits the existence of stationarity in at least a subset of the panels. Accounting for panel-specific autoregressive parameters, the test incorporates mean values and a time trend while excluding drift components. Cross-sectional means have been duly excised, and ADF regressions were implemented with one lag.

The statistical evidence, derived from various methodologies, robustly refutes the null hypothesis at a conventional significance level of p$<$ 0.01. This robust rejection provides persuasive evidence against the prevalence of unit roots within the panel dataset, intimating stationarity in at least one constituent panel.

Accordingly, the results from the Fisher-type unit-root test lend support to the stationarity of the ESG series across panels, challenging the premise of inherent non-stationarity. The presence of stationarity is a critical attribute that substantiates the econometric rigour of subsequent analyses. For additional verification, a Phillips–Perron unit root test was conducted, corroborating the initial findings. Refer to the Appendix for a more detailed elaboration.

\begin{figure}[htp!]
     \centering     
     \caption{Recession indicator, NBER. }
\includegraphics[width=0.9\textwidth]{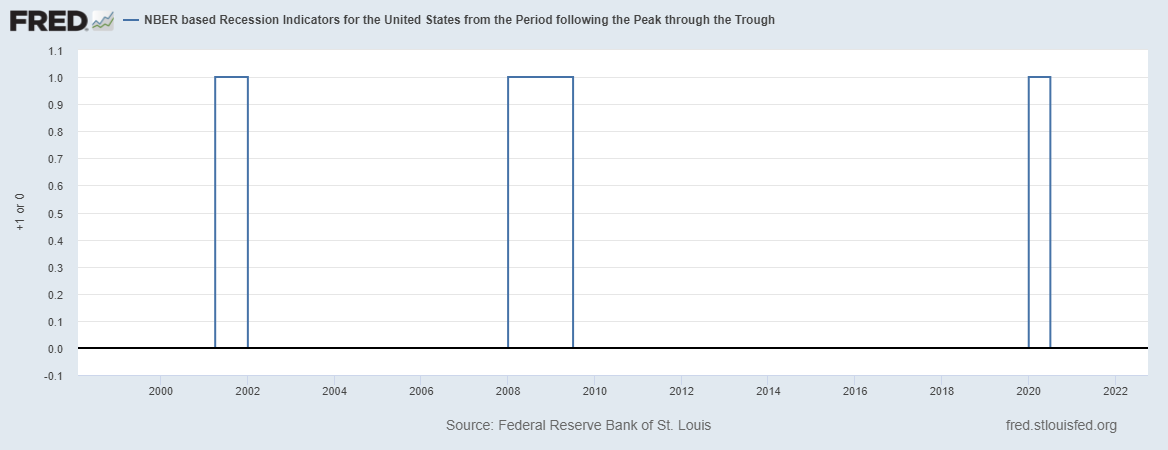}
     \label{fig:recessions}     
\end{figure}

Figure \ref{fig:recessions} illustrates the temporal distribution of recessionary periods within our dataset, as delineated by the National Bureau of Economic Research (NBER).

 %%%%%%%%%%%%%%%%%%%%%%%%%%%%%%%%%%%%%%% Plot mrecessions

\subsubsection{Hausman Test}
A pivotal aspect of the analysis is the selection of an appropriate panel data model to accurately capture the relationship between these factors and firm innovation. To this end, a Hausman Test was conducted to compare the fixed-effects and random-effects models in understanding the association between R\&D Index and various financial variables, including size, Tangibility, EBITDA, R\&D intensity, tax rate, dividend yield, Graham number, and recession indicators.

The Hausman test, a standard procedure in panel data analysis, tests the null hypothesis that the preferred model is RE against the alternative that FE is more appropriate. The test compares the coefficients of the FE and RE models; a significant difference indicates that the RE model's assumption of uncorrelated effects is violated, favouring the FE model. Our Hausman test yielded a Sargan-Hansen statistic of 7.780 with a Chi-squared distribution (7 degrees of freedom) and a p-value of 0.3524. This high p-value suggests that the differences in coefficients between the FE and RE models are not systematic, typically indicating that the RE model might be appropriate.

However, despite the Hausman test results, we argue that the FE model is more suitable for our analysis. The justification lies in the nature of our study and the specific characteristics of the variables involved. In assessing firm innovation, it is crucial to account for unobserved heterogeneity that could be correlated with the explanatory variables. The FE model is adept at controlling for such time-invariant heterogeneity within entities, effectively isolating the impact of explanatory variables on the dependent variable.

Particularly for the variable of interest, R\&D intensity, it's reasonable to assume that firm-specific, unobservable characteristics – such as management quality, corporate culture, or strategic priorities – could influence both the intensity of R\&D activities and other explanatory variables like 'total assets' or 'tangibility'. The FE model mitigates the bias that might arise from omitting these time-invariant characteristics, providing a more accurate estimation of the relationships of interest.

Furthermore, the consistency in the signs and significance of the coefficients across both FE and RE models supports the robustness of our model selection. The FE model's ability to control for unobserved, time-invariant heterogeneity, which is likely correlated with our explanatory variables, makes it a more appropriate choice for capturing the nuances of firm innovation dynamics.

In short, while the Hausman test statistically favours the RE model, theoretical considerations and the nature of our study strongly advocate for the use of the FE model. This approach ensures a more accurate and nuanced understanding of the factors driving firm innovation, crucial for both academic inquiry and practical implications.

\section{Econometric Modelling}
\label{sec:identification}
\subsection{Econometric Specifications}
Our econometric analysis is designed to investigate the determinants of R\&D Index, a measure of firm innovation. Common econometric approaches such as least squares methods and Tobit models are employed; however, these techniques face limitations due to their linearity assumptions and the absence of upper boundary considerations. To address these limitations, we employ Fractional Regression Models (FRM) and their panel data extensions (PFRM)\footnote{Please refer to Papke and Wooldridge (1996) for a detailed discussion on these models.}, which provide a robust and flexible framework. These models are particularly well-suited for variables like the R\&D Index, which are inherently bounded within a [0, 1] range. The fractional regression approach also accommodates the analysis of firms that may lack a rating, offering a comprehensive econometric structure for our investigation.

The primary models to be estimated are outlined as follows: The dependent variable, represented by $y_{it}$, denotes the R\&D Index. This variable is inherently constrained to the interval [0,1], reflecting the proportionate nature of R\&D intensity. In our modelling context, the index $i=1,\ldots, I$ corresponds to individual firms, and the index $t=1, \ldots, T$ denotes time.

\begin{flalign*} 
& \textit{1) Linear:} \\
& \textit{RnD}_{it} = \alpha_i + \beta_1 \xi_{it} + \beta_2 \tau_{it} + \beta_3 \eta_{it} + \beta_4 \rho_{it}  + \beta_5 \phi_{it} + \beta_6 \omega_{it} + \beta_7 \zeta_{it} + \beta_8 \theta_t + \epsilon_{it} \\\
\end{flalign*}

\begin{flalign*}
& \textit{2) Logit, linear-fractional:} \\
& \log\left(\frac{\textit{RnD}_{it}}{1-\textit{RnD}_{it}}\right) = \alpha_i + \gamma_1 \xi_{it} + \gamma_2 \tau_{it} + \gamma_3 \eta_{it}  + \gamma_4 \rho_{it} + \gamma_5 \phi_{it} + \gamma_6 \omega_{it}  + \gamma_7 \zeta_{it} + \gamma_8 \theta_t + \epsilon_{it}\\
\end{flalign*}

\begin{flalign*}
& \textit{3) Probit, linear-fractional:}\\
& \Phi^{-1}(\textit{RnD}_{it}) = \alpha_i + \delta_1 \xi_{it} + \delta_2 \tau_{it} + \delta_3 \eta_{it}  + \delta_4 \rho_{it} + \delta_5 \phi_{it} + \delta_6 \omega_{it}  + \delta_7 \zeta_{it} + \delta_8 \theta_t + \epsilon_{it} 
\end{flalign*}

Where the variables are defined as follows:
\begin{align*} \tiny
\xi_{it} & : \text{Size} \\
\tau_{it} & : \text{Tangibility} \\
\eta_{it} & : \text{EBITDA} \\
\rho_{it} & : \text{ESG} \\
\phi_{it} & : \text{Effective Tax Rate} \\
\omega_{it} & : \text{Dividend Yield} \\
\zeta_{it} & : \text{Grahams Number} \\
\theta_t & : \text{Recession}
\end{align*}

\subsection{Robustness Check: Incorporating Interaction Effects}
%\subsubsection{Extension: Addressing the Influence of Economic Downturns}
In an effort to bolster the robustness of our econometric analysis, we have considered the potential modulating impact of economic downturns on R\&D investment. To this end, we have constructed an interaction term that captures the interplay between the presence of a recessionary period and the R\&D Index of firms. This interaction term is then integrated into our regression framework to discern the extent to which economic crises may influence or interact with a firm's propensity to invest in innovation.

% Linear, with interaction effects:
\begin{flalign*}
& \textit{1) Linear:} \\
& \textit{RnD}_{it} = \alpha_i + \beta_1 \xi_{it} + \beta_2 \tau_{it} + \beta_3 \eta_{it} + \beta_4 \rho_{it} \\
& \phantom{\textit{RnD}_{it} =} + \beta_5 \phi_{it} + \beta_6 \omega_{it} + \beta_7 \zeta_{it} + \beta_8 \theta_t \\
& \phantom{\textit{RnD}_{it} =} + \beta_9 \theta_t \times \textit{RnD}_{it} + \epsilon_{it}
\end{flalign*}

\begin{flalign*}
& \textit{2) Logit, linear-fractional:} \\
& \log\left(\frac{\textit{RnD}_{it}}{1-\textit{RnD}_{it}}\right) = \alpha_i + \gamma_1 \xi_{it} + \gamma_2 \tau_{it} + \gamma_3 \eta_{it} \\
& \phantom{\log\left(\frac{\textit{RnD}_{it}}{1-\textit{RnD}_{it}}\right) =} + \gamma_4 \rho_{it} + \gamma_5 \phi_{it} + \gamma_6 \omega_{it} \\
& \phantom{\log\left(\frac{\textit{RnD}_{it}}{1-\textit{RnD}_{it}}\right) =} + \gamma_7 \zeta_{it} + \gamma_8 \theta_t + \gamma_9 \theta_t \times \textit{RnD}_{it} + \epsilon_{it}
\end{flalign*}

\begin{flalign*}
& \textit{3) Probit, linear-fractional:}\\
& \Phi^{-1}(\textit{RnD}_{it}) = \alpha_i + \delta_1 \xi_{it} + \delta_2 \tau_{it} + \delta_3 \eta_{it} \\
& \phantom{\Phi^{-1}(\textit{RnD}_{it}) =} + \delta_4 \rho_{it} + \delta_5 \phi_{it} + \delta_6 \omega_{it} \\
& \phantom{\Phi^{-1}(\textit{RnD}_{it}) =} + \delta_7 \zeta_{it} + \delta_8 \theta_t + \delta_9 \theta_t \times \textit{RnD}_{it} + \epsilon_{it}
\end{flalign*}

where $\theta_t \times \textit{RnD}_{it}$ is the Recession interaction term.

\subsection{Mitigating Concerns of Model Misspecification}

To address the potential confounding due to omitted variables and the simultaneity bias, our research employs dynamic panel data methods. These methods are instrumental in enhancing the precision and robustness of the estimated coefficients. Within the context of panel data analysis, we deploy fixed effects and random effects models to contend with unobserved heterogeneity.

The fixed effects approach is utilised to control for unobservable characteristics that are constant over time and specific to each entity, such as intrinsic firm attributes that may be correlated with the regressors. In contrast, the random effects framework operates under the assumption that unobserved heterogeneity is orthogonal to the regressors. Incorporating these effects is essential for the model to account for latent variables, thereby mitigating the risk of biased estimates.

\subsection{Robustness Check: Phillips–Perron Unit Root Test for Panel Data}

Complementing the ADF unit root test as described in Section \ref{sec:ADF}, we further scrutinise the stationarity of the panel data with the application of the Phillips–Perron unit root test, this time focusing on the R\&D Index variable. This test persists under the null hypothesis (H0) that each panel features a unit root, indicating a non-stationary series. The alternative hypothesis contends the existence of stationarity within at least one panel. Characteristic of this test is the consideration of panel-specific autoregressive parameters and mean values, with the exclusion of a time trend component. For this analysis, Newey–West lags were set to one.

The statistical outcomes, derived through various test methods, cohesively repudiate the null hypothesis at a conventional alpha level of p$<$ 0.01. Such a result provides substantial evidence refuting the universal presence of unit roots within the panels, thereby confirming the stationarity of at least one panel.

To summarise, the Phillips–Perron unit root test results align with those obtained from the ADF tests. The consistency of these results across different methodological approaches reinforces the argument against the existence of unit roots within our panel dataset.

\section{Empirical Hypotheses}
\label{sec:hypotheses}

In this study, we articulate hypotheses to examine the relationships between a variety of financial and operational variables and the R\&D Intensity of firms within the S\&P 500 index. We posit that total assets, profitability, and company size are positively correlated with R\&D Intensity, suggesting that larger and more profitable firms are likely to have a greater capacity and propensity for innovation. In contrast, Tangibility may exhibit a negative correlation with R\&D Intensity, as firms with a higher proportion of tangible assets could potentially allocate fewer resources to research and development activities.

%\subsection{Commentary on Hypotheses Validation}

In our empirical investigation, we set forth several hypotheses to explore the intricate relationships between various financial and operational variables and the R\&D Intensity of firms within the S\&P 500 index. The subsequent analysis, informed by rigorous regression models, has yielded insights that either validate or challenge our initial postulations.

\textit{Total Assets:} Our first hypothesis posited a positive relationship between total assets and R\&D Intensity, under the premise that firms with greater assets have more resources for innovation. The empirical evidence from our models corroborates this hypothesis. We observed a consistent positive association across different model specifications, affirming that firms with larger total assets are indeed more inclined to allocate significant funds towards R\&D initiatives. This finding aligns with the notion that resource abundance can facilitate greater investment in innovation activities.

\textit{EBITDA:} Contrary to hypothesis 2, which suggested a positive correlation between higher EBITDA and R\&D investment, our empirical results did not find a significant relationship. This outcome implies that a firm's earnings before interest, taxes, depreciation, and amortization may not be a primary driver of its R\&D spending. It could be indicative of the complex decision-making processes in firms where profitability does not straightforwardly translate into R\&D investment.

\textit{Tangibility:} Our analysis also delved into the relationship between tangibility and R\&D Intensity, hypothesizing an inverse relationship. The results partially support this hypothesis, showing a negative correlation in some model variants. This suggests that firms with a higher proportion of tangible assets might indeed prioritize physical asset investment over R\&D activities, possibly due to differing strategic focuses or the nature of their industry.

\textit{Size:} Interestingly, the hypothesis that larger firms would exhibit higher R\&D Intensity was not upheld. Instead, we found a consistent negative relationship between firm size and R\&D intensity, suggesting that as firms grow larger, the proportion of revenue allocated to R\&D decreases. This counterintuitive finding highlights the complexity of innovation dynamics in large corporations, possibly due to bureaucratic hurdles or diversified investment strategies.

\textit{Recession:} The impact of economic recessions on R\&D Intensity was hypothesized to be negative. Our study confirms this hypothesis, as recession periods were consistently associated with a downturn in R\&D activities. This result underscores the cautious approach firms adopt during economic uncertainties, often scaling back on discretionary expenditures like R\&D.

\textit{Tax Rate and Dividend Yield:} The hypotheses concerning the tax rate and dividend yield did not find strong empirical support in our models. These factors did not exhibit a significant impact on R\&D Intensity, suggesting that other variables might play more substantial roles in influencing R\&D investment decisions.

In conclusion, our empirical investigation offers a nuanced understanding of the factors that influence R\&D intensity in firms. While some hypotheses were validated, others were challenged or nuanced by the data, reflecting the multifaceted nature of corporate innovation strategies. 

The role of macroeconomic conditions, such as recessions and the Global Financial Crisis (GFC), is also scrutinised, with an anticipated adverse impact on R\&D Intensity. The hypothesis is that during economic downturns, firms may reduce investment in R\&D due to constrained budgets or uncertain futures. Additional factors, including the effective tax rate, dividend yield, and the Graham Number, are considered to delineate their intricate and potentially multi-dimensional relationships with a firm's R\&D Intensity. Together, these hypotheses are designed to offer a nuanced perspective on the factors influencing innovation within corporations, thereby enriching the dialogue around investment strategies and corporate innovation policies.

\begin{enumerate}
\item \textit{Total Assets:} 
  \begin{itemize}
  \item \textit{Hypothesis 1:} Firms with greater total assets may have more resources to allocate towards R\&D initiatives.
  \item \textit{Justification:} A positive association between total assets and R\&D Intensity could indicate a firm's capacity for sustaining significant investment in innovation.
  \end{itemize}

\item \textit{EBITDA:}
  \begin{itemize}
  \item \textit{Hypothesis 2:} Higher EBITDA may correlate with greater investment in R\&D.
  \item \textit{Justification:} Firms with higher earnings may have more financial flexibility to support R\&D activities, potentially leading to a higher R\&D Intensity.
  \end{itemize}

\item \textit{Tangibility:}
  \begin{itemize}
  \item \textit{Hypothesis 3:} Tangibility could inversely relate to R\&D Intensity.
  \item \textit{Justification:} Firms with a higher proportion of tangible assets may prioritize physical asset investment over R\&D, potentially resulting in lower R\&D Intensity.
  \end{itemize}

\item \textit{Size:}
  \begin{itemize}
  \item \textit{Hypothesis 4:} Larger firms may exhibit a higher R\&D Intensity.
  \item \textit{Justification:} Size may reflect the ability of firms to engage in substantial R\&D endeavors due to their larger resource base.
  \end{itemize}

\item \textit{Recession:}
  \begin{itemize}
  \item \textit{Hypothesis 5:} Economic recessions may adversely affect R\&D Intensity.
  \item \textit{Justification:} Recessions could lead firms to curtail discretionary spending, including R\&D investment, to conserve capital.
  \end{itemize}

\item \textit{Tax Rate:}
  \begin{itemize}
  \item \textit{Hypothesis 6:} The effective tax rate may influence R\&D Intensity through fiscal incentives or constraints.
  \item \textit{Justification:} Tax policies, including credits and incentives for R\&D, may affect a firm's propensity to invest in research and development.
  \end{itemize}

\item \textit{Dividend Yield:}
  \begin{itemize}
  \item \textit{Hypothesis 7:} Dividend yield may not be directly related to R\&D Intensity.
  \item \textit{Justification:} Dividend policies could be independent of a firm's investment in R\&D, reflecting different aspects of financial strategy.
  \end{itemize}
\end{enumerate}

While the Graham Number assesses the stock's intrinsic value, its connection to a firm's R\&D intensity remains to be explored. Thus, Graham Number's relationship with R\&D Intensity is uncertain and warrants investigation.

\section{Results}
\label{sec:results}
\subsection{Benchmark models}

\begin{table}[htp!]
    \centering
    \footnotesize
\begin{tabular}{l*{3}{c}}
\toprule
                    &\multicolumn{1}{c}{(1)}&\multicolumn{1}{c}{(2)}&\multicolumn{1}{c}{(3)}\\
                    &\multicolumn{1}{c}{Linear-FE}&\multicolumn{1}{c}{LOGIT}&\multicolumn{1}{c}{PROBIT}\\
                    &Coef./std.errors         &Coef./std.errors         &Coef./std.errors         \\
\midrule
size                &-0.0001849297${*}$  &-0.0019944911${*}$  &-0.0009417196${*}$  \\
                    &      (0.00)         &      (0.00)         &      (0.00)         \\
tangibility         &-0.0000102948         &-0.0001791661         &-0.0000872208         \\
                    &      (0.00)         &      (0.00)         &      (0.00)         \\
totalassets         &0.0000000000${+}$  &0.0000000000${*}$  &0.0000000000${*}$  \\
                    &      (0.00)         &      (0.00)         &      (0.00)         \\
ebitda              &0.0000000000         &0.0000000000         &0.0000000000         \\
                    &      (0.00)         &      (0.00)         &      (0.00)         \\
esg                 &0.0000025077         &0.0000381762         &0.0000173097         \\
                    &      (0.00)         &      (0.00)         &      (0.00)         \\
taxrate             &-0.0000000000         &-0.0000000011       &-0.0000000005         \\
                    &      (0.00)         &      (0.00)         &      (0.00)         \\
dividendYield       &-0.0000012344         &-0.0000150799       &-0.0000076470         \\
                    &      (0.00)         &      (0.00)         &      (0.00)         \\
grahamNumber        &0.0000000998         &0.0000004166         &0.0000002083         \\
                    &      (0.00)         &      (0.00)         &      (0.00)         \\
recession           &0.0000931721         &0.0017578652         &0.0008621640         \\
                    &      (0.00)         &      (0.00)         &      (0.00)         \\
\midrule
N                   &   45018             &   45016             &   45016         \\
AIC                 & -354699.2         & -223695.0         & -288069.3         \\
\bottomrule
\end{tabular}
    \caption{Benchmark models: OLS-FE,  Logit linear-fractional FE, Probit linear-fractional FE}
    \label{tab:main}
     \floatfoot{Dependent variable is RnD Index. Cluster-robust standard errors in parentheses. (+ 0.10 * 0.05 ** 0.01 *** 0.001)} 
\end{table}

%\subsection{Summary of benchmark models}
The regression results, summarized in Table~\ref{tab:main}, reveal several noteworthy patterns:

\begin{itemize}
    \item \textit{Size}: Consistently significant across all models, indicating a negative relationship with R\&D intensity. This might suggest that larger firms have a lower proportion of revenue allocated to R\&D activities.
    \item \textit{Tangibility}: Although not statistically significant in most models, it hints at a potential negative influence on R\&D intensity.
    \item \textit{Total Assets}: Shows a positive association with R\&D intensity in Logit and Probit models, suggesting that firms with larger asset bases tend to invest more in R\&D.
    \item \textit{EBITDA}: Not significantly associated with R\&D intensity, indicating that current earnings before interest, taxes, depreciation, and amortization may not be a strong predictor of R\&D spending.
    \item \textit{ESG}: The relationship with R\&D intensity is not conclusively determined due to mixed significance levels.
    \item \textit{Tax Rate and Dividend Yield}: Generally, these variables do not exhibit a significant impact on R\&D intensity.
    \item \textit{Graham Number}: Shows a limited relationship with R\&D intensity.
    \item \textit{Recession}: Emerges as a significant predictor, particularly in the Logit model, suggesting that economic downturns might influence R\&D investment decisions.
\end{itemize}

Based on the Akaike Information Criterion (AIC), the Logit linear-fractional FE model (Model 2) is identified as the most appropriate for our analysis, exhibiting the lowest AIC score among the three models. This model balances model complexity and goodness of fit, making it a robust choice for interpreting the determinants of R\&D intensity.

The findings from this study contribute to a nuanced understanding of how various factors influence firm innovation. The significance of firm size and the impact of economic conditions, as revealed in our models, offer valuable insights for policymakers and business strategists. Future research could further explore the interplay between these variables and R\&D intensity, considering additional factors such as industry-specific dynamics or geographical influences.

The results from our study present two interesting findings regarding the relationship between a firm's R\&D intensity and its total assets and size. Let's explore each of these findings with respect to the economic rationale based on the theory of the firm.

\subsubsection{Total Assets and R\&D Intensity}

%Positive Association:
The finding that larger total assets are associated with higher R\&D intensity in Logit and Probit models aligns with certain theoretical expectations. Firms with larger asset bases often have more resources at their disposal, which can be allocated to R\&D activities. This is consistent with the resource-based view of the firm, which posits that firms with more resources can undertake more extensive R\&D efforts.

%Economic Rationale: 
Larger asset bases often indicate a firm's capacity to invest in long-term projects, including R\&D. These firms may have more stable cash flows, a broader asset base to leverage for financing, and greater ability to absorb the risks associated with R\&D investments. Furthermore, larger firms might have more diversified operations, enabling them to invest in a wider range of R\&D activities.

\subsubsection{Firm Size and R\&D Intensity}

%Negative Relationship: 
The consistent significance of firm size across models, indicating a negative relationship with R\&D intensity, may initially seem counterintuitive. One might expect larger firms to spend more in absolute terms on R\&D. However, when considering R\&D intensity – R\&D spending as a proportion of revenue – this result can be explained.

%Economic Rationale:
Larger firms often experience economies of scale, which can lead to efficiency in operations and reduced per-unit costs. As firms grow, their focus may shift towards exploiting existing capabilities and market positions rather than investing proportionally more in R\&D. Additionally, larger firms might face bureaucratic inefficiencies or a lack of flexibility, which can impede innovative activities. This is in line with the theory of diminishing marginal returns, where the relative benefit of R\&D investment may decrease as the firm grows.

%Alternative Explanation: 
It's also possible that larger firms diversify their investment across various activities, not just R\&D. Hence, while their total R\&D spending might be high, it represents a smaller proportion of their overall revenue compared to smaller firms, which might focus more intensely on R\&D to achieve growth and competitive differentiation.

In summary, these findings suggest that while larger firms (in terms of total assets) tend to invest more in R\&D, the proportion of revenue they allocate to R\&D decreases as the firm's size increases. This is reflective of the complex relationship between firm size, resource availability, and strategic priorities in R\&D investment decisions.

\subsection{Robustness check: adding interaction effects}

To further enhance the analysis, we introduce an interaction term that combines the variable representing the occurrence of a crisis (Recession) with the dependent variable. We derive the interaction term and subsequently incorporate it into our regression analysis. The regression results are summarized in Table~\ref{tab:mains}.

\begin{table}[htp!]
    \centering
    \footnotesize
\begin{tabular}{l*{3}{c}}
\toprule
                    &\multicolumn{1}{c}{(1)}&\multicolumn{1}{c}{(2)}&\multicolumn{1}{c}{(3)}\\
                    &\multicolumn{1}{c}{Linear-FE}&\multicolumn{1}{c}{LOGIT}&\multicolumn{1}{c}{PROBIT}\\
                    &Coef./std.errors         &Coef./std.errors         &Coef./std.errors         \\
\midrule
size                &  -0.0001328*  &  -0.0013649** &  -0.0006295** \\
                    &      (0.00)   &      (0.00)   &      (0.00)   \\
tangibility         &   0.0000001   &  -0.0000511   &  -0.0000231   \\
                    &      (0.00)   &      (0.00)   &      (0.00)   \\
ebitda              &   0.0000000   &   0.0000000   &   0.0000000   \\
                    &      (0.00)   &      (0.00)   &      (0.00)   \\
esg                 &   0.0000024   &   0.0000363   &   0.0000164   \\
                    &      (0.00)   &      (0.00)   &      (0.00)   \\
taxrate             &   0.0000000   &   0.0000000   &   0.0000000   \\
                    &      (0.00)   &      (0.00)   &      (0.00)   \\
dividendYield       &  -0.0000000   &  -0.0000001   &  -0.0000001   \\
                    &      (0.00)   &      (0.00)   &      (0.00)   \\
grahamNumber        &   0.0000001   &   0.0000001   &   0.0000000   \\
                    &      (0.00)   &      (0.00)   &      (0.00)   \\
recession           &  -0.0408009***&  -0.5006269***&  -0.2507047***\\
                    &      (0.00)   &      (0.01)   &      (0.00)   \\
recession\_rnd       &   0.9294621***&  11.4185216***&   5.7177746***\\
                    &      (0.02)   &      (0.20)   &      (0.07)   \\
\midrule
N                   &   45018             &   45016             &   45016         \\
AIC                 &-357658.1   &-257040.5   &-323845.7   \\
\bottomrule
\end{tabular}
    \caption{Models with interaction effects: OLS-FE,  Logit linear-fractional FE, Probit linear-fractional FE}
    \label{tab:mains}
     \floatfoot{Dependent variable is RnD Index. Cluster-robust standard errors in parentheses. (+ 0.10 * 0.05 ** 0.01 *** 0.001)} 
\end{table}

The key findings from the interaction models are as follows:

\begin{itemize}
    \item The interaction term (recession\_rnd) is highly significant across all models, suggesting that the effect of a recession on R\&D intensity is substantial and varies depending on other firm-specific factors.
    \item The size of the firm continues to show a significant negative relationship with R\&D intensity, although the magnitude of this effect is somewhat reduced when considering the interaction with recession.
    \item The coefficient of the recession variable is negative and significant, indicating a general downturn in R\&D activities during recession periods. However, the positive and significant coefficient of the interaction term implies that this negative effect is moderated or reversed under certain conditions.
\end{itemize}

These results underscore the complexity of R\&D investment decisions, particularly during economic downturns. They suggest that the impact of a recession on firm innovation is not uniform but depends on other characteristics of the firm. This insight should be of interest to business strategists and policymakers, especially in formulating responses to economic crises.

\subsection{Robustness check: System GMM Models}

\begin{table}[htp!]
    \centering
    \footnotesize
\begin{tabular}{l*{3}{c}}
\toprule
                    &\multicolumn{1}{c}{(1)}&\multicolumn{1}{c}{(2)}&\multicolumn{1}{c}{(3)}\\
                    &\multicolumn{1}{c}{Linear-FE}&\multicolumn{1}{c}{LOGIT}&\multicolumn{1}{c}{PROBIT}\\
                    &Coef./std.errors         &Coef./std.errors         &Coef./std.errors         \\
\midrule
L.size              &   0.7372980    {***}&                     &                     \\
                    &      (0.00)         &                     &                     \\
size                &                     &  -0.0022114    {***}&  -0.0010804    {***}\\
                    &                     &      (0.00)         &      (0.00)         \\
tangibility         &  -0.0014724         &  -0.0002157    {*}  &  -0.0001116    {**} \\
                    &      (0.00)         &      (0.00)         &      (0.00)         \\
totalassets         &   0.0000000    {***}&   0.0000000    {***}&   0.0000000    {***}\\
                    &      (0.00)         &      (0.00)         &      (0.00)         \\
ebitda              &   0.0000000    {***}&  -0.0000000         &  -0.0000000         \\
                    &      (0.00)         &      (0.00)         &      (0.00)         \\
esg                 &   0.0042136    {***}&  -0.0000244         &  -0.0000114         \\
                    &      (0.00)         &      (0.00)         &      (0.00)         \\
taxrate             &   0.0000002         &  -0.0000000         &  -0.0000000         \\
                    &      (0.00)         &      (0.00)         &      (0.00)         \\
dividendYield       &  -0.0028658         &  -0.0000106         &  -0.0000056         \\
                    &      (0.00)         &      (0.00)         &      (0.00)         \\
grahamNumber        &   0.0000419    {**} &   0.0000000         &   0.0000000         \\
                    &      (0.00)         &      (0.00)         &      (0.00)         \\
recession           &  -0.0131605    {***}&   0.0007615    {***}&   0.0003871    {***}\\
                    &      (0.00)         &      (0.00)         &      (0.00)         \\
L.lyLOGIT           &                     &   0.7995575    {***}&                     \\
                    &                     &      (0.00)         &                     \\
L.lyPROBIT          &                     &                     &   0.7985424    {***}\\
                    &                     &                     &      (0.00)         \\
Constant            &   5.6159688    {***}&  -0.5664445    {***}&  -0.3189673    {***}\\
                    &      (0.07)         &      (0.01)         &      (0.00)         \\
\midrule
\end{tabular}
    \caption{Models with interaction effects: OLS-FE,  Logit linear-fractional FE, Probit linear-fractional FE}
    \label{tab:maingmm}
     \floatfoot{Dependent variable is RnD Index. Cluster-robust standard errors in parentheses. (+ 0.10 * 0.05 ** 0.01 *** 0.001)} 
\end{table}

Table \ref{tab:maingmm} presents the outcomes of an additional robustness check using System Generalized Method of Moments (GMM) Models. We adopt the Arellano and Bover (1995) and Blundell and Bond (1998) system GMM estimation approach for computing the regression coefficients. It is important to note that the time-dependent dummy variables' estimated coefficients are not included in the reported results.

For diagnostic purposes, we employ the AR(1) and AR(2) tests, as proposed by Arellano and Bond (1991), to detect any potential autocorrelation in the differenced data. Our models successfully clear both the AR(1) and AR(2) tests, thereby confirming the absence of autocorrelation concerns. Furthermore, the validity of the instruments used in our models is corroborated by passing the Sargan test for over identification restrictions, with a significance level of $p<0.01$. This underlines the robustness of our System GMM model estimations.

\begin{itemize}
    \item \textit{Linear-FE Model (1)}: The lagged size variable (L.size) shows a strong positive impact, indicating that past firm size significantly influences current R\&D Index. This implies a dynamic relationship where previous firm size conditions continue to affect current innovation strategies.
    \item \textit{LOGIT Model (2)}: In this model, current size (size) negatively impacts R\&D Index, while the lagged dependent variable from the LOGIT model (L.lyLOGIT) has a significant positive effect. This suggests that while larger firms tend to allocate a lower proportion of revenue to R\&D, past innovation intensity positively influences current innovation activities.
    \item \textit{PROBIT Model (3)}: Similar to the LOGIT model, current size negatively influences R\&D Index, and the lagged dependent variable from the PROBIT model (L.lyPROBIT) exerts a significant positive effect. This aligns with the LOGIT model's findings, reinforcing the importance of past innovation intensity in shaping current R\&D decisions.
\end{itemize}

The System GMM results validate the dynamic nature of firm innovation activities, highlighting the influence of both current and past firm characteristics on R\&D intensity. The consistency across models in the sign and significance of key variables like firm size and lagged dependent variables underscores the robustness of these findings.

%\subsection{Comparative Analysis of Model Outputs}

%\subsection{Results Summary}

\section{Discussion}
\label{sec:discussion}

This study aimed to examine the underlying factors influencing innovation, as measured by R \& D intensity, in S\&P 500 listed companies, offering an intricate landscape of their association with various financial and operational variables. Here, we extend the discussion to delve deeper into the complex web of associations revealed by our empirical analyses.

%\subsection{Insights for Policy}

The findings of this study offer valuable insights for policymakers, particularly in the realm of fostering corporate innovation and supporting economic growth. Based on our empirical analysis, several policy recommendations emerge:

 \textit{Enhancing Financial Resources for R\&D}: Given the positive correlation between total assets and R\&D intensity, policies that facilitate easier access to financial resources for firms could stimulate innovation. This could be achieved through tax incentives for R\&D investments, government grants, or subsidies specifically targeted at innovation activities.

 \textit{Support during Economic Downturns}: Our findings indicate a decline in R\&D intensity during recessions. To counteract this, counter-cyclical policies that provide additional support to firms during economic downturns could be beneficial. These may include tax breaks, increased government spending on research, or temporary relief from regulatory constraints to encourage continuous investment in R\&D.

 \textit{Encouraging SME Innovation}: The inverse relationship between firm size and R\&D intensity suggests that smaller firms might be more agile or inclined to invest a higher proportion of their revenue in innovation. Policies that support small and medium-sized enterprises (SMEs), such as easing capital access or providing innovation-specific funding, could thus be particularly effective in boosting overall industry innovation.

 \textit{Balancing Tangible and Intangible Investments}: The negative correlation between tangibility and R\&D intensity highlights the need for policies that help firms balance their investment in physical assets with intangible assets like R\&D. Educational initiatives and guidelines on effective investment strategies could be beneficial in this regard.

\textit{Fostering a Pro-Innovation Tax Environment}: Although the study found no significant direct impact of the tax rate on R\&D intensity, a conducive tax environment can indirectly support innovation. This includes tax policies that are favorable to corporate research and development, such as R\&D tax credits and deductions.

 \textit{Long-term Perspective in Policy Formulation}: Considering the complexity and time-lagged nature of innovation outcomes, it is crucial for policies to adopt a long-term perspective. This ensures sustained support for R\&D activities, even if immediate economic benefits are not evident.

Overall, the study underscores the importance of a multifaceted policy approach that considers various financial and economic factors influencing corporate innovation. By addressing these areas, policymakers can create an ecosystem that nurtures innovation, drives economic growth, and ensures long-term competitiveness.

%\subsection{Limitations and Future Research}

While this study provides valuable insights into the determinants of R\&D intensity in firms, it is important to acknowledge certain limitations that pave the way for future research opportunities.

One of the primary limitations lies in the scope of the data used. Our study focused on firms within the S\&P 500 index, which, while representative of large, publicly traded companies, may not capture the innovation dynamics of smaller firms or those in emerging markets. The innovation processes and financial constraints in these smaller or geographically diverse firms could differ significantly from those observed in the S\&P 500 firms.

Another limitation is the reliance on quantitative financial indicators to measure innovation. R\&D intensity, defined as R\&D expenditure relative to revenue, serves as a proxy for innovation but does not encompass all aspects of the innovation process. Qualitative factors such as organizational culture, management practices, and employee skills, which play a crucial role in innovation, were not captured in our analysis. Future studies could benefit from incorporating these qualitative aspects to provide a more holistic view of innovation within firms.

Additionally, the study primarily employed linear models to understand the relationships between financial indicators and R\&D intensity. However, these relationships could be non-linear or influenced by complex interactions between multiple variables. Future research might explore more sophisticated modelling techniques, such as non-linear models or machine learning algorithms, to uncover these potentially intricate relationships.

The time frame of the study also poses a limitation. Our analysis was based on data from a specific period, which may not account for long-term trends or cyclical variations in innovation investment. Longitudinal studies extending over a more extended period could provide deeper insights into how innovation strategies evolve over time and how they are influenced by economic cycles.

Furthermore, the impact of external macroeconomic factors, such as regulatory changes, technological advancements, and market competition, was not directly measured in this study. Future research could aim to integrate these external factors to better understand how they interact with internal financial indicators to influence a firm’s innovation activities.

In conclusion, while this study contributes to our understanding of the financial determinants of innovation in large corporations, it opens several avenues for future research. By addressing these limitations and exploring the outlined research opportunities, subsequent studies can build on our findings to further enrich the discourse on corporate innovation and its drivers.

\section{Conclusion}
\label{sec:conclusion}

%This research adds to the complex debate surrounding the factors that influence research and development (R\&D) expenditure and firm performance. 
Our findings point to a parallel narrative regarding the dynamics of R\&D investment choices. On one hand, there exists a positive correlation between higher total assets and increasing research and development (R\&D) intensity. This finding supports the resource-based view of the firm, which posits that the availability of resources is linked to the ability to engage in substantial R\&D endeavours. This implies that companies with significant assets have a strategic edge in enhancing their innovation trajectory by making continuous expenditures in research and development.

On the other hand, the observed inverse association between the size of a company and its level of research and development (R\&D) highlights the decreasing value of R\&D spending as companies grow. This phenomena points to operational and bureaucratic obstacles that may impede innovation in larger organisations, underscoring the intricacy of enhancing research and development efficiency in tandem with business expansion. The situation is further complicated by the notable interaction effect between recessionary periods and R\&D intensity, which demonstrates how economic downturns have varying impacts on enterprises' decisions regarding R\&D spending, depending on their distinct characteristics.

The implications of our investigation are three-fold. 

First, the results point to the significance of adopting a strategic approach to investing in research and development, taking into account the company's size, assets, and the current economic conditions. 

Second, our results indicate that policymakers and business strategists should contemplate customised actions to bolster research and development expenditure during periods of economic decline, acknowledging that the effects of recessions vary among different companies.

And third, the association between recessions and the level of research and development (R\&D) activity necessitates a nuanced comprehension of economic resilience and the factors that influence it. This suggests that organisations have the ability to utilise their unique characteristics in order to alleviate the adverse impacts of economic downturns on their innovation endeavours, given appropriate circumstances. This understanding is crucial for companies seeking to navigate periods of economic downturn without jeopardising their long-term capacity for innovation.

\newpage

\bibliographystyle{acm} % We choose the "plain" reference style
\bibliography{references}

\newpage
%%%%%%%%%%%%%%%%%%%%%%%%%%%%%%%%%%%%%%%%%%%%%%%%%%%%%%%%%%%%%%%%%%%%%%%%%5
\appendix
\section{VIF}
\begin{table}[htbp]
\caption{Multicollinearity Diagnostics}
\label{tab:vif}
\begin{tabular}{ll}
Variable      & VIF  \\ \hline
              &      \\
rndindex      & 1.00 \\
size          & 1.47 \\
tangibility   & 1.00 \\
totalassets   & 1.47 \\
ebitda        & 1.34 \\
esg           & 1.00 \\
taxrate       & 1.00 \\
dividendyield & 1.00 \\
grahamnumber  & 1.01
\end{tabular}
\end{table}

\section{Robustness check - System GMM Models with interaction effects}

\begin{table}[htp!]
    \centering
    \footnotesize
\begin{tabular}{l*{3}{c}}
\toprule
                    &\multicolumn{1}{c}{(1)}&\multicolumn{1}{c}{(2)}&\multicolumn{1}{c}{(3)}\\
                    &\multicolumn{1}{c}{Linear-FE}&\multicolumn{1}{c}{LOGIT}&\multicolumn{1}{c}{PROBIT}\\
                    &Coef./std.errors         &Coef./std.errors         &Coef./std.errors         \\
\midrule
L.size              &   0.7372040    {***}&                     &                     \\
                    &      (0.00)         &                     &                     \\
size                &                     &  -0.0029763    {***}&  -0.0014492    {***}\\
                    &                     &      (0.00)         &      (0.00)         \\
tangibility         &  -0.0014062         &  -0.0000439         &  -0.0000224         \\
                    &      (0.00)         &      (0.00)         &      (0.00)         \\
totalassets         &   0.0000000    {***}&   0.0000000    {***}&   0.0000000    {***}\\
                    &      (0.00)         &      (0.00)         &      (0.00)         \\
ebitda              &   0.0000000    {***}&  -0.0000000         &  -0.0000000         \\
                    &      (0.00)         &      (0.00)         &      (0.00)         \\
esg                 &   0.0042067    {***}&   0.0000124         &   0.0000067         \\
                    &      (0.00)         &      (0.00)         &      (0.00)         \\
taxrate             &   0.0000002         &   0.0000000         &   0.0000000         \\
                    &      (0.00)         &      (0.00)         &      (0.00)         \\
dividendYield       &  -0.0028451         &  -0.0000075         &  -0.0000035         \\
                    &      (0.00)         &      (0.00)         &      (0.00)         \\
grahamNumber        &   0.0000419    {**} &  -0.0000001         &  -0.0000000         \\
                    &      (0.00)         &      (0.00)         &      (0.00)         \\
recession           &   0.2731793    {***}&  -0.2917617    {***}&  -0.1522085    {***}\\
                    &      (0.08)         &      (0.00)         &      (0.00)         \\
recession\_rnd       &  -6.5154175    {***}&   6.6490566    {***}&   3.4684796    {***}\\
                    &      (1.77)         &      (0.04)         &      (0.02)         \\
L.lyLOGIT           &                     &   0.5174614    {***}&                     \\
                    &                     &      (0.00)         &                     \\
L.lyPROBIT          &                     &                     &   0.4937611    {***}\\
                    &                     &                     &      (0.00)         \\
Constant            &   5.6184471    {***}&  -1.4207844    {***}&  -0.8321571    {***}\\
                    &      (0.07)         &      (0.01)         &      (0.00)         \\
\midrule
\end{tabular}
    \caption{System GMM Models with interaction effects.}
    \label{tab:maingmms}
     \floatfoot{Dependent variable is RnD Index. Cluster-robust standard errors in parentheses. (+ 0.10 * 0.05 ** 0.01 *** 0.001)} 
\end{table}

Table \ref{tab:maingmms} presents the outcomes of an additional robustness check using System Generalized Method of Moments (GMM) Models.

\textit{Lagged Size (Linear-FE Model)}: A notable result is the positive and highly significant coefficient of lagged firm size (L.size). This indicates that the previous period's size of a firm has a lasting and positive impact on its current R\&D intensity, suggesting a cumulative effect of firm size on innovation efforts over time.

 \textit{Current Size (LOGIT and PROBIT Models)}: Both models reveal a significant negative relationship between the current size of the firm and R\&D intensity. This finding is consistent across models, reinforcing the notion that larger firms, in terms of their current size, tend to allocate a smaller proportion of their revenue to R\&D activities.

 \textit{Total Assets and EBITDA (Linear-FE Model)}: Total assets and EBITDA show a significant positive relationship with R\&D intensity. This result underscores the importance of financial strength and earnings in facilitating a firm's investment in R\&D.

 \textit{ESG (Linear-FE Model)}: The environmental, social, and governance (ESG) factor displays a significant positive correlation with R\&D intensity. This indicates that firms with better ESG performance are likely to invest more in R\&D.

 \textit{Recession and Interaction Term}: The inclusion of the recession variable and its interaction with R\&D intensity (recession\_rnd) in all models reveals significant coefficients. The Linear-FE model shows a significant negative impact of the interaction term, while the LOGIT and PROBIT models demonstrate a significant positive impact. This highlights the complex and varied influence of economic downturns on R\&D investment decisions.

 \textit{Lagged Dependent Variables (LOGIT and PROBIT Models)}: The significant coefficients of the lagged dependent variables (L.lyLOGIT and L.lyPROBIT) emphasize the role of past innovation intensity in shaping a firm's current R\&D activities.

These significant results from the System GMM models with interaction effects enhance our understanding of the factors influencing R\&D intensity in firms. The findings highlight the dynamic nature of innovation investment decisions, influenced by a firm's size, financial health, ESG performance, economic conditions, and past innovation activities.

\section{Explanation of Variables and Dataset}

\textit{Variables Used in the Analysis:}
\begin{table}[ht]
\centering
\begin{tabular}{|l|l|}
    \hline
    \textit{Variable} & \textit{Description} \\
    \hline
    ESG Score & Firm's exposure to governance, social, and environmental risks. \\
    Total Assets & Net income divided by return on assets. \\
    EBITDA & Earnings before interest, taxes, depreciation, and amortization. \\
    Recession & NBER Recession Indicator. \\
    GFC & NBER Indicator for the 2007--2008 financial crisis. \\
    Size & Natural logarithm of total assets. \\
    Tangibility & Tangible assets divided by total assets. \\
    Tax Rate & Provision for income taxes divided by income before tax. \\
    Dividend Yield & Dividend per share divided by price. \\
    Graham Number & \(\sqrt{22.5 \times \text{EPS} \times \text{BVPS}}\). \\
    R\&D Intensity & R\&D expenditure divided by revenue. \\
    \hline
\end{tabular}
\end{table}

\textit{Dataset Overview:}
\begin{itemize}
    \item Time Span: 1998q2 to 2023q2.
    \item Scope: All listed companies in the SP500 index.
    \item Focus: Firms listed in the United States.
    \item Exclusions: European exchanges due to reporting disparities.
    \item R\&D Intensity: Scaled to (0, 1) - corresponds to 0-100\%
    \item Sample Size: 506 firms, 100 periods, 45,386 observations.
\end{itemize}

\newpage
\section{List of abbreviations}

\begin{table}[htp!]
\centering \footnotesize
\begin{tabular}{ll}
Abbreviation & Explanation                                           \\ \toprule
GDP          & Gross domestic product                                \\
USRECQ       & US Recession Indicator (NBER)                \\
GFC          & Great Financial Crisis (2007–2008)                               \\
SP500           & Standard and Poor's 500 index                              \\
NBER         & National Bureau of Economic Research                  \\
EBITDA         & Earnings before interest, taxes, depreciation and amortization \\
R\&D         & Research and Development                              \\
SMEs         & Small and medium-sized enterprises                    \\
        
\end{tabular}
\caption{List of abbreviations used in the study}
\end{table}

\end{document}

%% file: preamble.tex
\makeatletter
\def\ps@pprintTitle{%
 \let\@oddhead\@empty
 \let\@evenhead\@empty
 \def\@oddfoot{}%
 \let\@evenfoot\@oddfoot}
\makeatother
\usepackage[capposition=top]{floatrow}
\usepackage{stackengine}
\usepackage[mathscr]{euscript}
\usepackage{subcaption}
\usepackage[colorlinks=false]{hyperref}
\usepackage{longtable}
\usepackage{booktabs}
 \usepackage{amssymb}
\usepackage{microtype} 
\hypersetup{pdfborder=0 0 0}
\usepackage{amsmath,amsthm,amssymb,bm}
\usepackage{booktabs}
\usepackage{threeparttable}

\usepackage{amssymb}
\usepackage[overload]{empheq}
\usepackage{latexsym,dsfont,textcomp}

\usepackage{commath,mathtools,mathrsfs}
\usepackage{parskip}
\usepackage{epstopdf,pdfpages,graphicx}
\usepackage{array}
\usepackage{babelbib}
\usepackage[english]{babel}

\providecommand{\keywords}[1]
{
  \small	
  \textbf{\textit{Keywords: }} #1
}

\providecommand{\JEL}[1]
{
  \small	
  \textbf{\textit{JEL Classification: }} #1
}

\graphicspath{ {./images/} }
{
      \theoremstyle{plain}
      
  }

\newtheorem{innercustomgeneric}{\customgenericname}
\providecommand{\customgenericname}{}
\newcommand{\newcustomtheorem}[2]{%
  \newenvironment{#1}[1]
  {%
   \renewcommand\customgenericname{#2}%
   \renewcommand\theinnercustomgeneric{##1}%
   \innercustomgeneric
  }
  {\endinnercustomgeneric}
}

\newcustomtheorem{axiom}{Axiom}
\newcustomtheorem{hypo}{Hypothesis}

\usepackage{lipsum} % filler text

\usepackage[nameinlink,capitalize]{cleveref}

\usepackage{listings}
\usepackage{color}

\lstdefinelanguage{Stata}{
    morekeywords=[1]{regress, reg, summarize, sum, display, di, generate, gen, bysort, use, import, delimited, predict, quietly, probit, margins, test},
    morekeywords=[2]{aggregate, array, boolean, break, byte, case, catch, class, colvector, complex, const, continue, default, delegate, delete, do, double, else, eltypedef, end, enum, explicit, export, external, float, for, friend, function, global, goto, if, inline, int, local, long, mata, matrix, namespace, new, numeric, NULL, operator, orgtypedef, pointer, polymorphic, pragma, private, protected, public, quad, real, return, rowvector, scalar, short, signed, static, strL, string, struct, super, switch, template, this, throw, transmorphic, try, typedef, typename, union, unsigned, using, vector, version, virtual, void, volatile, while,},
    morekeywords=[3]{forvalues, foreach, set},
    morekeywords=[4]{bofd, Cdhms, Chms, Clock, clock, Cmdyhms, Cofc, cofC, Cofd, cofd, daily, date, day, dhms, dofb, dofC, dofc, dofh, dofm, dofq, dofw, dofy, dow, doy, halfyear, halfyearly, hh, hhC, hms, hofd, hours, mdy, mdyhms, minutes, mm, mmC, mofd, month, monthly, msofhours, msofminutes, msofseconds, qofd, quarter, quarterly, seconds, ss, ssC, tC, tc, td, th, tm, tq, tw, week, weekly, wofd, year, yearly, yh, ym, yofd, yq, yw,},
    morekeywords=[5]{abs, ceil, cloglog, comb, digamma, exp, expm1, floor, int, invcloglog, invlogit, ln, ln1m, ln, ln1p, ln, lnfactorial, lngamma, log, log10, log1m, log1p, logit, max, min, mod, reldif, round, sign, sqrt, sum, trigamma, trunc,},
    morekeywords=[6]{cholesky, coleqnumb, colnfreeparms, colnumb, colsof, corr, det, diag, diag0cnt, el, get, hadamard, I, inv, invsym, issymmetric, J, matmissing, matuniform, mreldif, nullmat, roweqnumb, rownfreeparms, rownumb, rowsof, sweep, trace, vec, vecdiag, },
    morekeywords=[7]{autocode, byteorder, c, _caller, chop, abs, clip, cond, e, fileexists, fileread, filereaderror, filewrite, float, fmtwidth, has_eprop, inlist, inrange, irecode, matrix, maxbyte, maxdouble, maxfloat, maxint, maxlong, mi, minbyte, mindouble, minfloat, minint, minlong, missing, r, recode, replay, return, s, scalar, smallestdouble,},
    morekeywords=[8]{rbeta, rbinomial, rcauchy, rchi2, rexponential, rgamma, rhypergeometric, rigaussian, rlaplace, rlogistic, rnbinomial, rnormal, rpoisson, rt, runiform, runiformint, rweibull, rweibullph,},
    morekeywords=[9]{tin, twithin,},
    morekeywords=[10]{betaden, binomial, binomialp, binomialtail, binormal, cauchy, cauchyden, cauchytail, chi2, chi2den, chi2tail, dgammapda, dgammapdada, dgammapdadx, dgammapdx, dgammapdxdx, dunnettprob, exponential, exponentialden, exponentialtail, F, Fden, Ftail, gammaden, gammap, gammaptail, hypergeometric, hypergeometricp, ibeta, ibetatail, igaussian, igaussianden, igaussiantail, invbinomial, invbinomialtail, invcauchy, invcauchytail, invchi2, invchi2tail, invdunnettprob, invexponential, invexponentialtail, invF, invFtail, invgammap, invgammaptail, invibeta, invibetatail, invigaussian, invigaussiantail, invlaplace, invlaplacetail, invlogistic, invlogistictail, invnbinomial, invnbinomialtail, invnchi2, invnF, invnFtail, invnibeta, invnormal, invnt, invnttail, invpoisson, invpoissontail, invt, invttail, invtukeyprob, invweibull, invweibullph, invweibullphtail, invweibulltail, laplace, laplaceden, laplacetail, lncauchyden, lnigammaden, lnigaussianden, lniwishartden, lnlaplaceden, lnmvnormalden, lnnormal, lnnormalden, lnwishartden, logistic, logisticden, logistictail, nbetaden, nbinomial, nbinomialp, nbinomialtail, nchi2, nchi2den, nchi2tail, nF, nFden, nFtail, nibeta, normal, normalden, npnchi2, npnF, npnt, nt, ntden, nttail, poisson, poissonp, poissontail, t, tden, ttail, tukeyprob, weibull, weibullden, weibullph, weibullphden, weibullphtail, weibulltail,},
    morekeywords=[11]{abbrev, char, collatorlocale, collatorversion, indexnot, plural, plural, real, regexm, regexr, regexs, soundex, soundex_nara, strcat, strdup, string, strofreal, string, strofreal, stritrim, strlen, strlower, strltrim, strmatch, strofreal, strofreal, strpos, strproper, strreverse, strrpos, strrtrim, strtoname, strtrim, strupper, subinstr, subinword, substr, tobytes, uchar, udstrlen, udsubstr, uisdigit, uisletter, ustrcompare, ustrcompareex, ustrfix, ustrfrom, ustrinvalidcnt, ustrleft, ustrlen, ustrlower, ustrltrim, ustrnormalize, ustrpos, ustrregexm, ustrregexra, ustrregexrf, ustrregexs, ustrreverse, ustrright, ustrrpos, ustrrtrim, ustrsortkey, ustrsortkeyex, ustrtitle, ustrto, ustrtohex, ustrtoname, ustrtrim, ustrunescape, ustrupper, ustrword, ustrwordcount, usubinstr, usubstr, word, wordbreaklocale, worcount,},
    morekeywords=[12]{acos, acosh, asin, asinh, atan, atanh, cos, cosh, sin, sinh, tan, tanh,},
    morecomment=[l]{//},
    morecomment=[s]{/*}{*/},
    morestring=[b]{`}{'},
    morestring=[b]",
    morestring=[d]",
    sensitive=true,
}

\definecolor{codeblue}{rgb}{0.29296875, 0.51953125, 0.68359375}
\definecolor{codegreen}{rgb}{0.47265625, 0.62890625, 0.40234375}
\definecolor{codegray}{rgb}{0.95703125, 0.95703125, 0.95703125}
\definecolor{codecrimson}{rgb}{0.87109375,0.3984375,0.3984375}